\documentclass[aps,prd,twocolumn,nofootinbib]{revtex4-1}
\usepackage{epsfig}
\usepackage[colorlinks,linkcolor=blue,anchorcolor=blue,citecolor=blue,urlcolor=blue,breaklinks=true]{hyperref}
\usepackage{hyperref}
\usepackage{graphics}
\usepackage{color}
\usepackage{slashed}
\usepackage{makecell}
\usepackage{subfigure}

\begin{document}
\author{Shu-Sheng Xu$^{1}$}\email[]{Email: xuss@njupt.edu.cn}
\address{$^{1}$ School of Science, Nanjing University of Posts and Telecommunications, Nanjing 210023, China}

\title{Phase structures of neutral dense quark matter and application to strange stars}
\begin{abstract}
In the contact interaction model, the quark propagator has only one solution, namely, the chiral symmetry breaking solution, at vanishing temperature and density with the case of physical quark mass. Inspired by Y. Jiang and Z.-F. Cui~\cite{PhysRevD.85.034031,cui2013wigner}, we introduce 2+1 flavors quark condensates feedback onto the coupling strength, and find the Wigner solution appears in some region, which enables us to tackle chiral phase transition as two-phase coexistences. At finite chemical potential, we analyze the chiral phase transition in the conditions of electric charge neutrality and $\beta$ equilibrium. The four chemical potentials, $\mu_u$, $\mu_d$, $\mu_s$ and $\mu_e$, are constrained by three conditions, so that there is one independent variable remains: we choose the average quark chemical potential as the free variable. All quark masses and number densities suffer discontinuities at the phase transition point. The strange quarks appear after the phase transition since the system needs more energy to produce a $d$-quark than an $s$-quark. Take the EOS as an input, the TOV equations are solved numerically, we show that the mass-radius relation is sensitive to the EOSs. The maximum mass of strange quark stars is not susceptible to the parameter $\Lambda_q$ we introduced.

\bigskip

\noindent Key-words: chiral phase transition, dense quark matter, quark stars

\bigskip

\noindent PACS Number(s): 11.30.Rd, 25.75.Nq, 12.38.Mh, 12.39.-x

\end{abstract}
\maketitle
\section{Introduction}\label{intro}
It is well-known that hadrons with strangeness are unstable, they could decay into lighter hadrons made of non-strange quarks through the weak interaction. The hypothesis of stable strange nuclei is first proposed by A. Bodmer in Ref.~\cite{PhysRevD.4.1601}. Afterward, E. Witten proposed stable dense quark matter containing strange quarks based on the assumption that the Fermi momentum of the system exceeds the strange quark mass in the dense quark matter, therefore the dense quark matter favors some non-strange quarks transform into strange quarks~\cite{PhysRevD.30.272}. There are plenty of studies on strange quark stars from that time on~\cite{PhysRevD.73.114016,PhysRevLett.96.041101,PhysRevD.43.627,tikekar2007relativistic,rahaman2014new,kalam2013relativistic,PhysRevD.100.043015}. Especially, after the gravitational waves are detected from the merging of compact stars, people are interested in exploring the inner structure of compact stars~\cite{PhysRevLett.126.162702,wiktorowicz2017strange,husain2021hybrid,kuerban2019gw}.

The properties of dense QCD play a key role in the structure of compact stars, but lattice regularized QCD can not deal with such systems since the technical difficulty, that is ``sign problem''. Various effective models are inevitably used to study the phase structures of dense QCD systems, such as the quasi-particle model~\cite{PhysRevD.82.014023,PhysRevD.84.094004,szabo2003quasiparticle,PhysRevC.79.055207,PhysRevC.64.055201,PhysRevC.69.035210,PhysRevC.61.045203}, quark-meson model~\cite{SCHAEFER2005479,PhysRevD.88.074006,PhysRevD.76.074023,KAMIKADO20131044,PhysRevD.97.034022,TETRADIS200393}, Nambu-Jona-Lasinio~(NJL) model~\cite{endrHodi2019magnetized,WAKAYAMA2019548,lu2015critical,khunjua2019qcd,PhysRevD.94.014026,cui2014wigner,cui2013wigner,PhysRevD.99.016018}, and some effective models of Dyson-Schwinger equations~(DSEs)~\cite{PhysRevD.102.034027,shi2014locate,jiang2013chiral,PhysRevD.90.034022,FISCHER2011438,FISCHER2014774,PhysRevD.93.036006,PhysRevD.94.094030,PhysRevD.101.054032,shi2020chiral,GAO2021136584,PhysRevD.91.014024,PhysRevD.93.034013}. Based on these model studies, people believe that there are colorful phase structures in dense QCD matter. Most of these model studies find there is a critical endpoint~(CEP) in the phase diagram. The chiral phase transition happens at low temperature and high density, while the low density and high-temperature region is crossover. In terms of cold dense QCD, DSEs confront a difficulty that the quark propagator has poles at high baryon chemical potential. NJL model also has a drawback that there is only one solution in the non-chiral limit, and the pressure of the system is discontinuous at the chiral phase transition point, therefore the equation of state~(EOS) is incomplete. In order to resolve this problem, the quark condensates feedback approach is proposed in the contact interaction model~\cite{PhysRevD.85.034031,cui2013wigner}. In this approach, the coupling strength depends on the quark condensates, and the Wigner solution appears for some parameter choice.

In this work, we will borrow this idea and extend it to 2+1 flavors to study dense QCD phase structure and strange quark stars.
This paper is organized as follows. We give a basic introduction to the quark gap equation and the quark condensate feedback approach in the 2+1 flavor case in Sec.~\ref{contactmodel}, the solutions of the quark propagator are shown and related parameters are fixed by pion and kaon observables. In Sec.~\ref{phasediagram}, the dense QCD phase structures are studied in conditions of electric charge neutrality and $\beta$ equilibrium. Using the EOS as an input, the mass-radius relation is given in Sec.~\ref{strangestar}. And finally, we summarize our studies in Sec.~\ref{sum}.
\section{Quark propagators in vaccum}\label{contactmodel}
\begin{figure}[!h]
\centering
\includegraphics[width=0.45\textwidth]{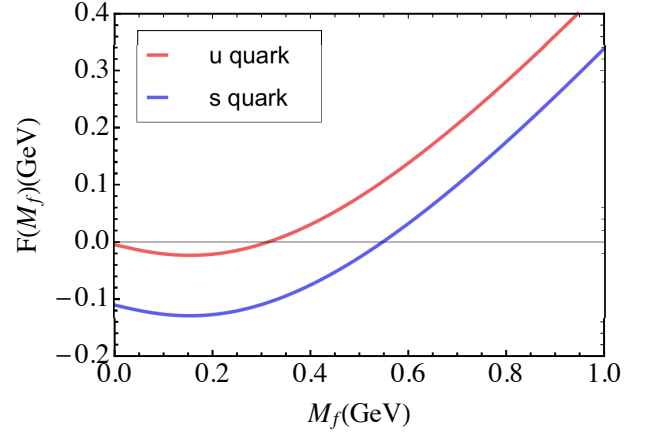}
\caption{The function $F(M_f)$ for u and s quarks.}\label{fig1}
\end{figure}

The quark propagator plays a central role in many issues, it satisfies the Dyson-Schwinger equation,
\begin{eqnarray}
S_f^{-1}(p) = S_{f0}^{-1}(p) + g^2 \int\frac{d^4q}{(2\pi)^4} D_{\mu\nu}(p-q) \gamma_\mu t^a S_f(p) \Gamma_\nu^a(p,q).
\label{quarkDSE}
\end{eqnarray}
where $S_{f0}(p)$ is the bare quark propagator with flavor $f$, $S_f(p)$ is the dressed quark propagator, $t^a~(a=1,2,\cdots,8)$ are half of Gell-Mann matrices, $\Gamma_\nu^a(p,q)$ is the amputated quark-gluon vertex, $D_{\mu\nu}(p-q)$ is the dressed gluon propagator. It is an exact equation, which is derived from QCD generating functional. The quark propagator (two-point Green function) is related to the quark-gluon vertex (three-point Green function), and the three-point Green function is related to higher-point Green functions, so that there are infinite coupled integral equations. Thus, specific truncation for DSEs is necessary. We will use rainbow truncation, which is simple and preserves chiral symmetry,
\begin{eqnarray}
\Gamma_\nu^a(p,q) \longrightarrow \gamma_\nu t^a,
\label{rainbow}
\end{eqnarray}
so that widely used in hadron physics as well as thermal and dense QCD.
The quark DSE is closed if dressed gluon propagator is specified. We use the contact interaction model in this work~\cite{PhysRevC.75.015201}, that is
\begin{equation}
g^2 D_{\mu\nu}(p-q) = \frac{1}{M_G^2}\delta_{\mu\nu}.
\label{CImodel}
\end{equation}
The quark DSE turns to be
\begin{equation}
S_f^{-1}(p) = S_{f0}^{-1}(p) + \frac{4}{3M_G^2}\int^\Lambda\frac{d^4q}{(2\pi)^4} \gamma_\mu S_f(p) \gamma_\mu.
\label{quarkDSE2}
\end{equation}
The integration in Eq.~(\ref{quarkDSE2}) with this model is divergent, the 3-dimension cut-off is performed in this work~\cite{RevModPhys.64.649}.
The inverse of quark propagator has the general structure
\begin{equation}
S_f^{-1}(p) = i\gamma\cdot p A_f(p^2) + B_f(p^2), \label{quarkprop}
\end{equation}
where $A_f(p^2)$ and $B_f(p^2)$ are two scalar functions.
Substituting Eqs.~(\ref{rainbow}), (\ref{CImodel}) and (\ref{quarkprop}) into Eq.~(\ref{quarkDSE}), one can obtain $A_f(p^2)=1$, $B_f(p^2)=M_f$ is a constant mass, which satisfies
\begin{eqnarray}
M_f &=& m_f + \frac{4}{3M_G^2} \int^\Lambda \mathrm{Tr}_D[S_f(p)]
\nonumber\\
&=& m_f + \frac{4}{3M_G^2} \int^\Lambda \frac{d^4q}{(2\pi)^4} \frac{4M_f}{q^2+M_f^2}
\nonumber\\
&=& m_f + \frac{2M_f}{3M_G^2\pi^2} \bigg[ \Lambda \sqrt{\Lambda^2 + M_f^2}
\nonumber\\
&&\hspace*{2cm} - M_f^2 \ln\left( \frac{\Lambda}{M_f} + \sqrt{1+ \frac{\Lambda^2}{M_f^2}} \right) \bigg].
\end{eqnarray}
There are four parameters in this model, namely $m_u=m_d$, $m_s$, $M_G$ and $\Lambda$, which are fitted by pion mass $m_\pi$, pion decay constant $f_\pi$, kaon mass $m_K$, and light quark condensate $\langle \bar\psi\psi$~\cite{Zyla:2020zbs,PhysRevD.87.034514}. The meson masses $m_\pi$, $m_K$ are the solutions of their Bethe-Salpeter equations,
\begin{eqnarray}
\Gamma_\pi(P) &=& -\frac{4}{3M_G^2} \int^\Lambda \frac{d^4q}{(2\pi)^4} \gamma_\mu S_u(q_+) \Gamma_\pi(P) S_u(q_-)\gamma_\mu,
\\
\Gamma_K(P) &=& -\frac{4}{3M_G^2} \int^\Lambda \frac{d^4q}{(2\pi)^4} \gamma_\mu S_u(q_+) \Gamma_K(P) S_s(q_-)\gamma_\mu.
\end{eqnarray}
where $q_+=q+P/2$, $q_-=q-P/2$, and $P^2=-m_\pi^2$ for pion, $P^2=-m_K^2$ for kaon. The pion decay constant in 3 momentum cut-off is~\cite{RevModPhys.64.649}
\begin{equation}
f_\pi^2 = N_c M_u^2 \int^\Lambda \frac{d^3\vec{q}}{(2\pi)^3} \frac{1}{(\vec{q}^2+M^2)^{3/2}}.
\end{equation}
  The parameters and corresponding observables are listed in Table.~\ref{table1}.
\begin{table*}
\centering
\begin{tabular}{|c|c|c|c|c|c|c|c|c|c|c|}
\hline
$m_u$   &$m_s$      &$M_G$      &$\Lambda$  &$M_u$  &$M_s$  &$m_\pi$    &$m_K$      &$f_\pi$    &$-\langle \bar u u\rangle^{\frac{1}{3}}$  &$-\langle \bar s s\rangle^{\frac{1}{3}}$  \\
\hline
4.3 &110    &189    &799    &314.6  &547.5    &139.3    &498.2    &93.3     &292.2    &327.6    \\
\hline
\end{tabular}
\caption{Model parameters and observables in vacuum~(all quantities in MeV).}
\label{table1}
\end{table*}
To analysis the solution of quark DSE, it is helpful to define a function
\begin{equation}
F(M_f) = M_f - m_f - \frac{2M_f}{3M_G^2\pi^2} \mathcal{C}(M_f,\Lambda),
\end{equation}
with
\begin{equation}
\mathcal{C}(M_f,\Lambda) = \Lambda \sqrt{\Lambda^2 + M_f^2} - M_f^2 \ln\left( \frac{\Lambda}{M_f} + \sqrt{1+ \frac{\Lambda^2}{M_f^2}} \right).
\end{equation}
One can plot the function $F(M_f)$ for $u$- and $s$-quarks, as shown in Fig.~\ref{fig1}. We can see that both functions have only one zero point, which implies there is only one solution in vacuum.

\begin{figure}[bp]
\centering
\subfigure[$\Lambda_q=0.4~\mathrm{GeV}$.]{
\includegraphics[width=0.22\textwidth]{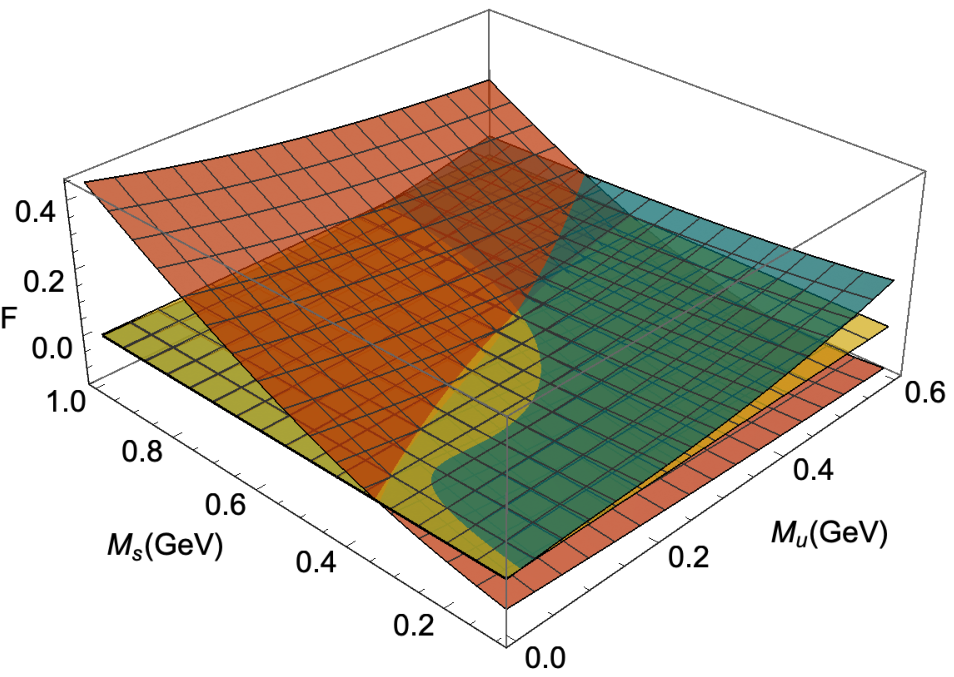}
}
\subfigure[$\Lambda_q=0.381~\mathrm{GeV}$.]{
\includegraphics[width=0.22\textwidth]{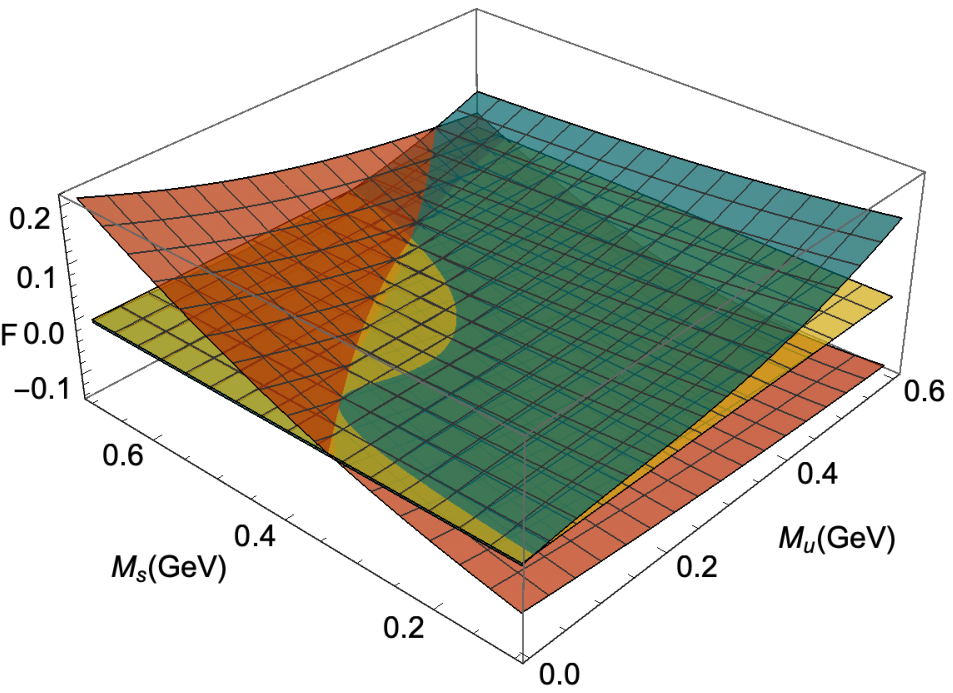}
}
\subfigure[$\Lambda_q=0.35~\mathrm{GeV}$.]{
\includegraphics[width=0.22\textwidth]{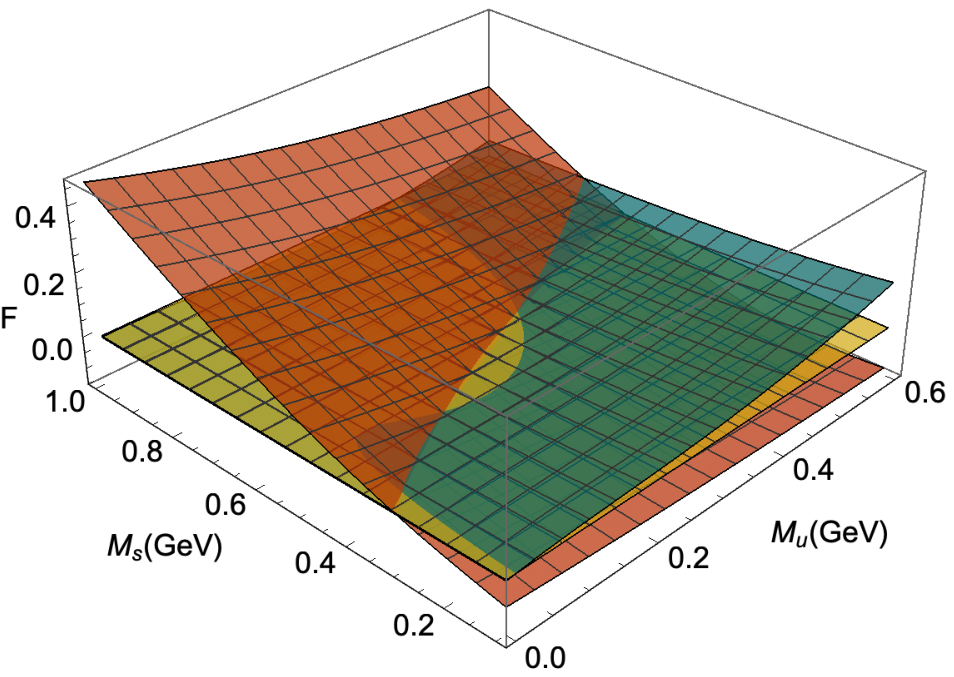}
}
\subfigure[$\Lambda_q=0.30~\mathrm{GeV}$.]{
\includegraphics[width=0.22\textwidth]{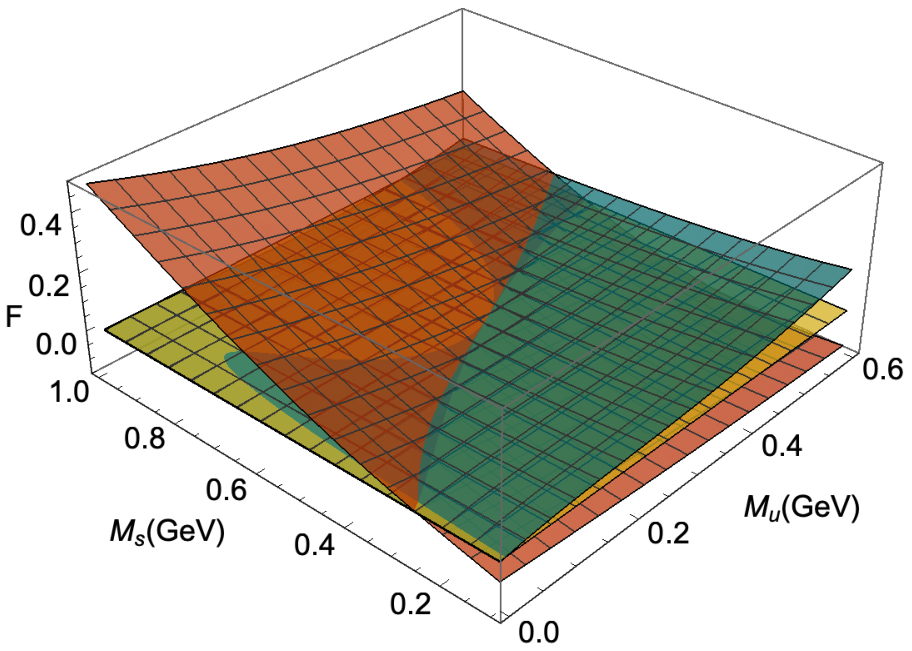}
}
\subfigure[$\Lambda_q=0.15~\mathrm{GeV}$.]{
\includegraphics[width=0.22\textwidth]{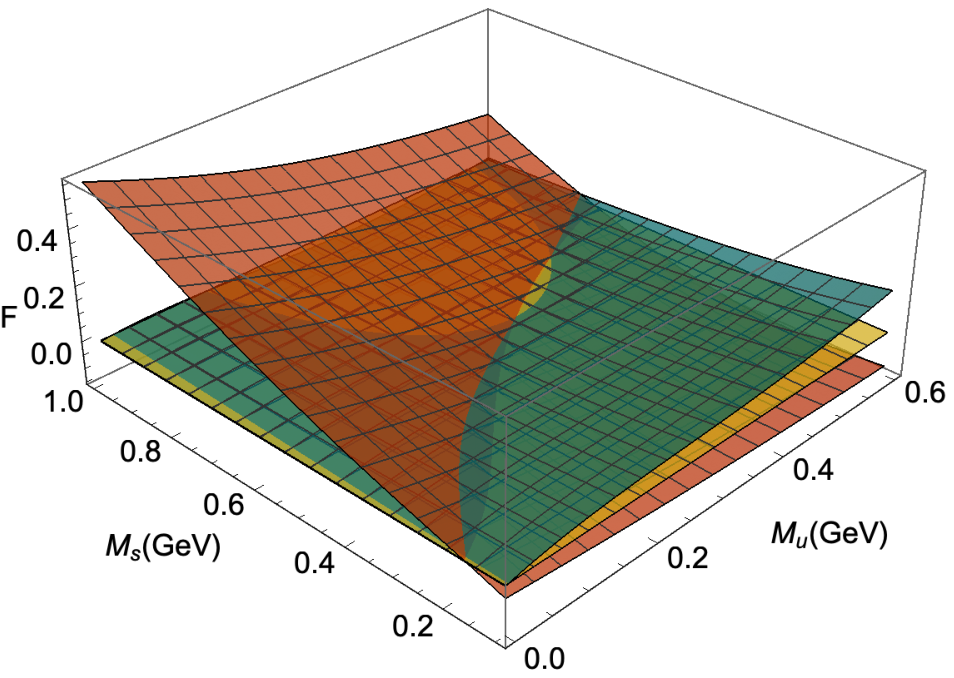}
}
\caption{The $F_u$ and $F_s$ as functions of $M_u$ and $M_s$, the green curved surface is $F_u$, the red one is $F_s$, and the yellow one is $F=0$. The intersections of three surfaces are the solutions of the quark DSEs.}
\label{fig2}
\end{figure}

According to the gluon DSE, the inverse of gluon propagator includes two parts, namely pure Yang-Mills terms and quark-loop term.
The effective interaction between quarks, which is related to the dressed gluon propagator, should depend on the feedback of the quark-loop. Therefore, the effective interaction would be different for different solutions. Inspired by Ref.~\cite{PhysRevD.85.034031}, we use the interaction
\begin{eqnarray}
g^2D_{\mu\nu}(k)=\delta_{\mu\nu}\frac{1}{M_{\mathrm{eff}}^2} = \delta_{\mu\nu}\left(\frac{1}{M_G^{\prime 2}} - \sum_{f=u,d,s}\frac{1}{M_G^{\prime 2}}\frac{\langle \bar ff\rangle}{\Lambda_f^3}\right),
\label{effective}
\end{eqnarray}
where
\begin{equation}
\langle\bar ff\rangle = \frac{N_c M_f}{2\pi^2}\mathcal{C}(M_f,\Lambda).
\end{equation}
Although there are three energy scales $\Lambda_f$, we would reduce parameters, namely, use the same value $\Lambda_u=\Lambda_d=\Lambda_s=\Lambda_q$, in this work. The modified interaction introduces two new parameters, $M_G^\prime$ and $\Lambda_q$. As the first step, we fix $M_{\mathrm{eff}}=M_G$ in the Nambu solution and discuss the dependence of the parameters. It implies
\begin{equation}
M_G^{\prime 2} = M_G^2\left( 1 - \sum_{f=u,d,s} \frac{\langle\bar ff\rangle}{\Lambda_q^3}  \right).
\end{equation}
 The quark DSEs of 3 flavors are coupled to each other, we can define two functions
\begin{eqnarray}
F_u(M_u,M_s) &=& M_u - m_u - \frac{2M_u}{3M_\mathrm{eff}^2\pi^2} \mathcal{C}(M_u,\Lambda),
\nonumber\\
F_s(M_u,M_s) &=& M_s - m_s - \frac{2M_s}{3M_\mathrm{eff}^2\pi^2} \mathcal{C}(M_s,\Lambda).
\end{eqnarray}
with $\frac{1}{M_{\mathrm{eff}}^2}$ is presented in Eq.~(\ref{effective}). The solution locates at conditions of $F_u(M_u,M_s)=0$ and $F_s(M_u,M_s)=0$. We plot them in Fig.~\ref{fig2}, the green curved surface is $F_u(M_u,M_s)$ and the red is $F_s(M_u,M_s)$. We can see that there is only one solution if $\Lambda_q>0.381$~GeV. When $\Lambda_q$ decreases to $0.381$~GeV, the second solution begins to appear. The third solution arises if $\Lambda_q<0.381~\mathrm{GeV}$, but $\Lambda_q$ should be restricted between 0.30~GeV and 0.381~GeV if we require the largest mass of the solution is the same as the original Nambu solution's mass. Up to now, the model parameters are fixed except $\Lambda_q$, we will use $\Lambda_q=0.38$~GeV to study the properties of neutral dense quark matter in the next section. Thereafter, we will discuss the EOS and mass-radius relation dependence on $\Lambda_q$.

\section{The chiral phase structure in neutral dense quark matter}\label{phasediagram}
We now turn our attention to the QCD system at finite chemical potential. The general structure of the inverse of quark propagator is~\cite{PhysRevD.85.034031}
\begin{eqnarray}
S_f^{-1}(p,\mu_f) = i\gamma\cdot p A_f(p,\mu_f) + B_f(p,\mu_f) - \gamma_4 C_f(p,\mu_f).
\label{quarkprop2}
\end{eqnarray}
Inserting Eq.~(\ref{quarkprop2}) into Eq.~(\ref{quarkDSE2}), one can easily find that $A_f(p,\mu_f)=1$, $B_f(p,\mu_f)=M_f$, and $C_f(p,\mu_f)=\mu_f^\ast$ are constant quantities, where $M_f$ and $\mu_f^\ast$ can be regarded as effective mass and chemical potential respectively. They satisfy
\begin{eqnarray}
M_f &=& m_f + \frac{4}{3M^2_{\mathrm{eff}}} \int^\Lambda \frac{d^4q}{(2\pi)^4} \frac{4M_f}{q^2+M_f^2-{\mu_f^\ast}^2 + 2i\mu_f^\ast q_4}
\nonumber\\
&=& m_f + \frac{4}{3M^2_{\mathrm{eff}}} \int^\Lambda \frac{d^3\vec{q}}{(2\pi)^4} \int_{-\infty}^{\infty} dq_4 \frac{4M_f}{q_4^2 + 2i\mu_f^\ast q_4 + E_{qMf}^2 -{\mu_f^\ast}^2},
\\
\mu_f^\ast &=&\mu_f - \frac{4}{3M^2_{\mathrm{eff}}} \int^\Lambda \frac{d^4q}{(2\pi)^4} \frac{2(iq_4 - \mu_f^\ast)}{q_4^2 + E_{qMf}^2 - {\mu_f^\ast}^2 + 2i\mu_f^\ast q_4}
\nonumber\\
&=&\mu_f - \frac{8}{3M^2_{\mathrm{eff}}} \int^\Lambda \frac{d^3\vec{q}}{(2\pi)^4} \int_{-\infty}^{\infty} dq_4 \frac{iq_4-\mu_f^\ast}{q_4^2 + 2i\mu_f^\ast q_4 +E^2_{qMf}-{\mu_f^\ast}^2},
\end{eqnarray}
with
\begin{equation}
E_{qMf} = \sqrt{\vec{q}^2 + M_f^2}.
\end{equation}
After some algebraic derivations, one finally obtains
\begin{eqnarray}
M_f &=& m_f + \frac{2M_f}{3\pi^2M^2_{\mathrm{eff}}} \mathcal{D}(\mu_f^\ast,M_f),    \label{DSE1}
\\
\mu^\ast_f &=& \mu_f - \frac{2}{3\pi^2M^2_{\mathrm{eff}}} \left( {\mu_f^\ast}^2 - M_f^2 \right)^{3/2} \theta(\mu_f^\ast-M_f),   \label{DSE2}
\\
\frac{1}{M^2_{\mathrm{eff}}} &=& \frac{1}{M_G^{\prime 2}} + \sum_{f=u,d,s} \frac{1}{M_G^{\prime 2}\Lambda_q^3} \frac{N_c M_f}{2\pi^2} \mathcal{D}(\mu_f^\ast,M_f),    \label{DSE3}
\end{eqnarray}
with
\begin{eqnarray}
&&\mathcal{D}(\mu_f^\ast, M_f) = \left[ \Lambda\sqrt{\Lambda^2 + M_f^2} - M_f^2\ln\left( \frac{\Lambda}{M_f} + \sqrt{1+\frac{\Lambda^2}{M_f^2}} \right) \right]
\nonumber\\
&&- \theta(\mu_f^\ast - M_f) \left[ \mu_f^\ast\sqrt{{\mu_f^\ast}^2-M_f^2} - M_f^2\ln\left( \frac{\sqrt{{\mu_f^\ast}^2-M_f^2}}{M_f} + \frac{\mu_f^\ast}{M_f} \right) \right].
\nonumber\\
\end{eqnarray}

There is no difficulty in principle to solve the coupled Eqs.~(\ref{DSE1}), (\ref{DSE2}) and (\ref{DSE3}) iteratively. Every quantity, such as effective mass, quark number densities, are functions of three different chemical potentials. In this work, we will focus on the phase structures in conditions of electric charge neutrality and $\beta$ equilibrium. The $\beta$ equilibrium requires
\begin{figure}[t]
\centering
\includegraphics[width=0.45\textwidth]{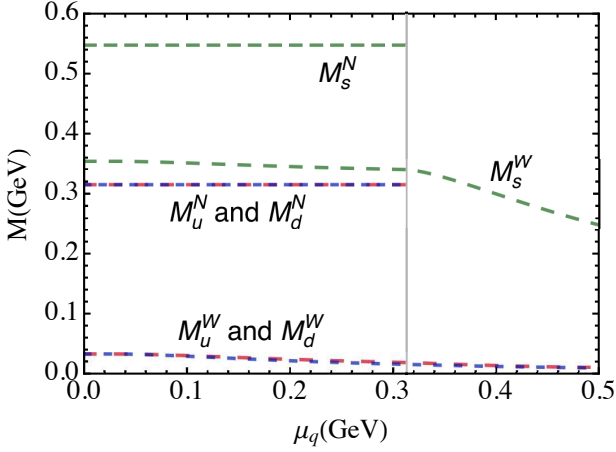}
\caption{Effective masses of quarks for both Nambu and Wigner solutions, the superscript `N' means Nambu solution, `W' means Wigner solution.\label{fig3}}
\end{figure}
\begin{figure}[b]
\centering
\includegraphics[width=0.45\textwidth]{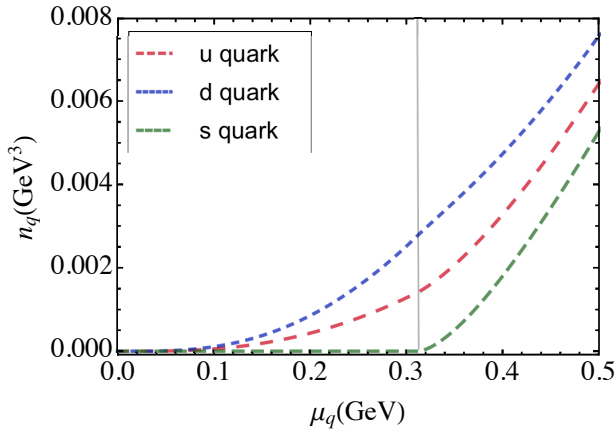}
\caption{Quark number densities as functions of $\mu_q$.\label{fig4}}
\end{figure}
\begin{eqnarray}
\mu_d&=&\mu_u+\mu_e,
\\
\mu_s&=&\mu_d.
\end{eqnarray}
The condition of electric charge neutrality is
\begin{equation}
\frac{2}{3}n_u - \frac{1}{3}n_d -\frac{1}{3}n_s - n_e = 0,
\end{equation}
where $n_u, n_d, n_s$ and $n_e$ are particle number densities for $u$-, $d$-, $s$-quarks and electron. We define the mean quark chemical potential by
\begin{equation}
\mu_q = \frac{\mu_u n_u + \mu_d n_d + \mu_s n_s}{n_u + n_d + n_s}.
\end{equation}

The four independent chemical potentials are constrained by three conditions, there is only one independent chemical potential, we are free to choose $\mu_q$ as that one. Effective masses of quarks varying with $\mu_q$ are displayed in Fig.~\ref{fig3}. We can see from Fig.~\ref{fig3} that the Nambu solutions keep to be constants at $\mu_q<315$~MeV, and disappear thereafter, while the Wigner solution exists on the whole region of quark chemical potential. In principle, the physical solution has the maximum pressure, and the bag constant, namely the pressure difference between Nambu and Wigner solutions in vacuum, is often introduced as a parameter in the NJL-like models. We assume the chiral phase transition happens at the location of Nambu solution disappears, that is $\mu_q^c=315$~MeV.

The quark number densities dependencies with $\mu_q$ are displayed in Fig.~\ref{fig4}. The densities of $u$- and $d$-quark for the Nambu solution appear at $314.6$~MeV and are very small compared to the densities of the Wigner solution. The densities suffer a discontinuity since the nature chooses Wigner solution at $\mu_q>\mu_q^c$. We can see that the $s$-quark appears very early because the quark condensates are smaller in the Wigner phase, which in turn affects the effective interaction strength. The nature also chooses the Wigner solution for $s$ quark, its effective mass $M^W_s(\mu_q=315\mathrm{MeV})=339$~MeV is lower than its effective chemical potential $\mu_s^\ast(\mu_q=315\mathrm{MeV})=367$~MeV.

\begin{figure}[h]
\centering
\includegraphics[width=0.5\textwidth]{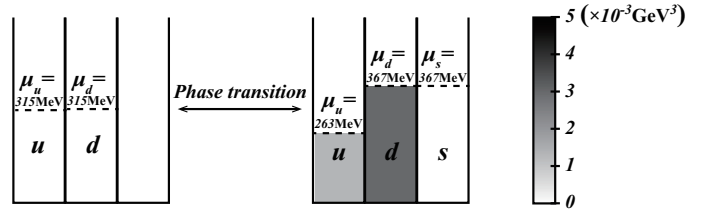}
\caption{The schematic illustration of the change of chemical potentials and quark number densities in the chiral phase transition~(The quark number densities are represented by darkness).}\label{fig5}
\end{figure}

Based on our calculations, the densities of up and down quarks are very small, and the strange quark does not exist in the Nambu phase. After the chiral phase transition, nature favors the Wigner solution, all quarks have lower effective masses, densities suffer discontinuities. The strange quark appears simultaneously. The transfer of chemical potentials is illustrated in Fig.~\ref{fig5}.
\section{Application to strange quark stars}\label{strangestar}
\begin{figure}[h]
\centering
\includegraphics[width=0.4\textwidth]{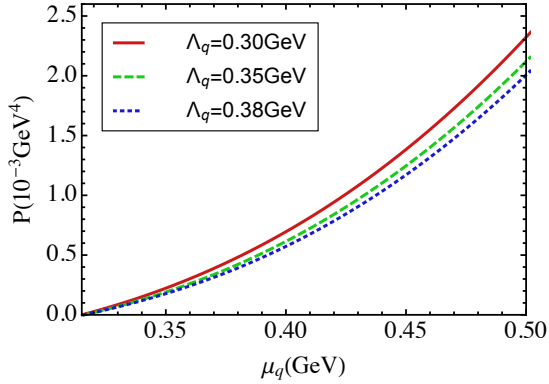}
\caption{Pressure as function of $\mu_q$ with three parameters.}\label{figp}
\end{figure}

\begin{figure}[h]
\centering
\includegraphics[width=0.4\textwidth]{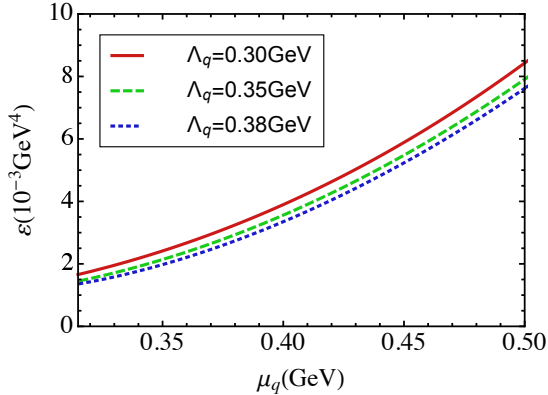}
\caption{Energy density as function of $\mu_q$ with three parameters.}\label{fige}
\end{figure}

\begin{figure}[h]
\centering
\includegraphics[width=0.4\textwidth]{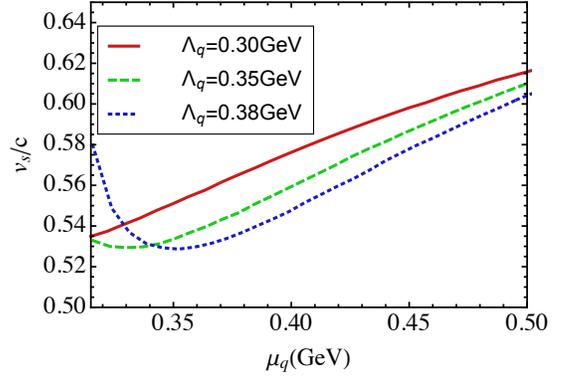}
\caption{Sound velocity as function of $\mu_q$ with three parameters.}\label{figs}
\end{figure}

\begin{figure}[h]
	\begin{minipage}[t]{0.4\textwidth}
		\centering
		\includegraphics[width=0.9\textwidth]{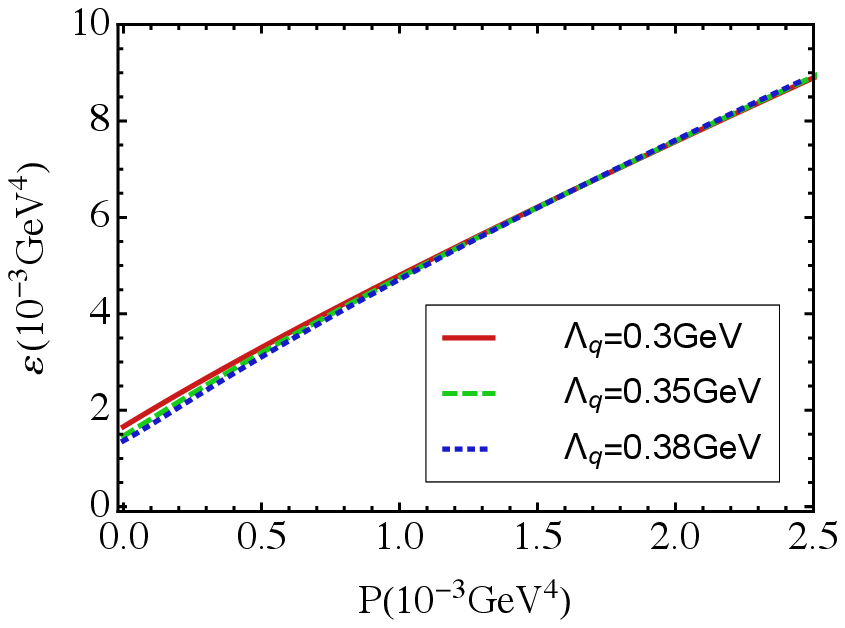}
		\caption{The EOS for three different parameters.\label{fig6}}
	\end{minipage}
	\qquad
	\begin{minipage}[t]{0.4\textwidth}
		\centering
		\includegraphics[width=0.9\textwidth]{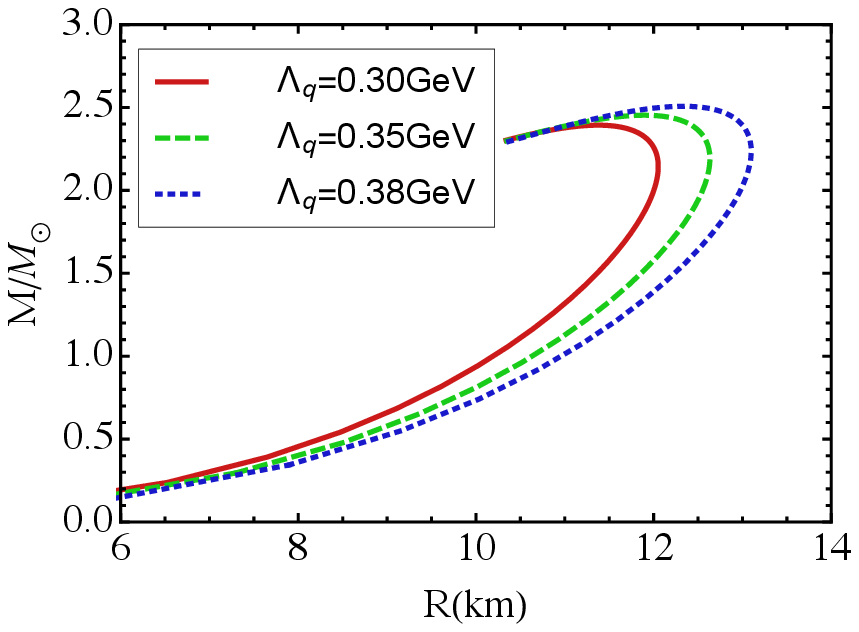}
		\caption{Mass-radius relation of strange quark stars.\label{fig7}}
	\end{minipage}
\end{figure}

The EOS of cold and dense quark matter plays a significant role in studying the structure of compact stars. We plot the EOS in Fig.~\ref{fig6}, which is based on the phase properties of dense quark matter in conditions of electric charge neutrality and $\beta$ equilibrium. The pressure is calculated by~\cite{zong2008model}
\begin{eqnarray}
P(\mu_q) &=& \int_0^{\mu_q} \Big(n_u(\mu_q^\prime) d\mu_u(\mu_q^\prime) + n_d(\mu_q^\prime) d\mu_d(\mu_q^\prime)
\nonumber\\
&&\hspace*{8mm} + n_s(\mu_q^\prime) d\mu_s(\mu_q^\prime) +n_e(\mu_q^\prime) d\mu_e(\mu_q^\prime)\Big).
\end{eqnarray}
It has relation with energy density
\begin{equation}
\varepsilon(\mu_q) = -P(\mu_q) + \sum_{i} \mu_i(\mu_q) n_i(\mu_q).
\end{equation}
The pressure, energy density and sound velocity $v_s=\sqrt{\frac{\partial P}{\partial \varepsilon}}$ as functions of $\mu_q$ are shown in Fig.~\ref{figp}, Fig.~\ref{fige} and Fig.~\ref{figs}. The main range of sound velocity of strange quark matter is $0.52<v_s/c<0.62$, which meets the requirement of relativity.

In Fig.~\ref{fig6}, we can see that the EOSs tend to coincident in the large pressure region, while they have little difference at low pressure. It implies the EOS is not too parameter-dependent in this model.

\begin{figure}[h]
\centering
\includegraphics[width=0.4\textwidth]{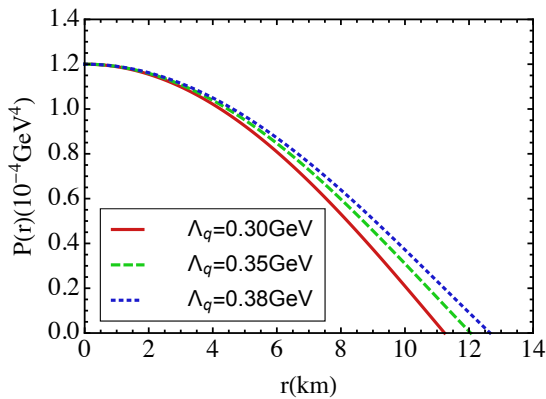}
\caption{The radius dependence of pressure at $P(0)=1.2\times 10^{-4}~\mathrm{GeV}^4$ for $\Lambda_q=(0.3, 0.35, 0.38)~$GeV.}\label{fig11}
\end{figure}

The relation between mass and radius of compact stars is governed by Tolman-Oppenheimer-Volkoff (TOV) equation,
\begin{eqnarray}
\frac{dP(r)}{dr} &=& -\frac{G(\varepsilon + P)(M + 4\pi r^3 P)}{r(r-2GM)},\label{TOV1}
\\
\frac{dM(r)}{dr} &=& 4\pi r^2 \varepsilon,\label{TOV2}
\end{eqnarray}
where the natural unit is used, $\hbar=1=c$. Take the EOS as input into Eqs.~(\ref{TOV1}) and (\ref{TOV2}), one can solve the TOV equation numerically, the mass-radius relation is illustrated in Fig.~\ref{fig7}. We can see that the curves are apparently separate from each other, which indicates that the mass-radius relations are very sensitive to the EOSs. When we take a look at the maximum mass of the strange quark star, they are (2.39, 2.45, 2.51)$M_\odot$ for $\Lambda_q=(0.30,0.35,0.38)~$GeV respectively. We find that the maximum mass is not sensitive to the parameter $\Lambda_q$. In order to show that details of the radius dependence of pressure, the $P(r)$ is plotted in Fig.~\ref{fig11}, we take the central pressure $P(0)=1.2\times 10^{-4}~\mathrm{GeV}^4$ as an example for three cases of parameters $\Lambda_q=(0.3, 0.35, 0.38)~$GeV.

\section{Summary and discussion}\label{sum}
We generalize the two-flavor contact interaction model to the three-flavor case, the model parameters are fitted by pion and kaon observables. Analyzing the solution of quark DSE, there is only one solution, Nambu solution or chiral symmetry breaking solution, in vacuum. Based on the ideas from Ref.~\cite{PhysRevD.85.034031,cui2013wigner}, we introduce a feedback term from quark condensates to the coupling strength. The Wigner solution appears at some region of $\Lambda_q$. We mainly discuss $\Lambda_q\in (0.3,0.381)$~GeV, since the Nambu and Wigner solution coexists and the largest mass equals the original Nambu one.

After fixing the model parameters, we study the chiral phase transition in the conditions of electric charge neutrality and $\beta$ equilibrium. During the phase transition, all quark masses and number densities suffer discontinuities. The strange quark appears since nature favors the lower effective mass and $\mu_s^\ast>M_s$. At the phase transition point, the two phases coexist. The $u$-quark has smaller chemical potential in the chiral symmetry partially restored phase, while $d$-quark has larger value of chemical potential. Therefore, the system would need more energy to produce a $d$-quark than a $s$-quark if the density of the $s$-quark is very small.

EOSs of dense quark matter with neutrality is shown for three different $\Lambda_q$, they are almost coincident in the large pressure region, and have little difference at low pressure. Take EOSs as inputs to the TOV equations, the mass-radius relation for strange quark stars are drawn, which are very sensitive to the EOSs. The maximum mass of strange quark stars is not susceptible to the parameter $\Lambda_q$, its value reaches $2.39M_\odot$ or even $2.51M_\odot$.

The strange quark have two effects on the EOSs, one is the quark condensate feedback, another is the nonzero strange quark number density diminishes the energy density. Comparing with the two flavor case of NJL model, the maximum mass of strange star is larger than that of neutron star~\cite{Xu:2021alh}. The strange stars are investigated with various models, such as quasi-particle model and NJL model~\cite{PhysRevD.99.043001,PhysRevD.101.063023}, the mass-radius relation are qualitatively consistent with these studies. In Ref.~\cite{Miao:2021nuq}, the authors analysis the strange stars using observational data of the GW170817 and GW190425 binary mergers, the inferred mass-radius relation is close to our curve with $\Lambda_q=0.38$~GeV.

\acknowledgments
This work is supported in part by the National Natural Science Foundation of China (under Grant No. 11905107), the National Natural Science Foundation of Jiangsu Province of China (under Grant No. BK20190721), Natural Science Foundation of the Jiangsu Higher Education Institutions of China (under Grant No. 19KJB140016), Nanjing University of Posts and Telecommunications Science Foundation (under grant No. NY129032), Innovation Program of Jiangsu Province.

\bibliography{references}
\end{document}